\documentclass[preprint,prb,showpacs,preprintnumbers,amsmath,amssymb,citeautoscript]{revtex4}

\usepackage{dcolumn}
\usepackage{bm}
\usepackage{graphicx,subfigure,endnotes,float}

\newcommand{\rv}{{\bf r}}
\newcommand{\pv}{{\bf p}}
\newcommand{\Tstar}{T^*}

\begin{document}

\title{Negative thermal expansion in single-component systems with
isotropic interactions}

\author{Mikael C. Rechtsman$^1$} \author{Frank H. Stillinger$^{2}$}
\author{Salvatore Torquato$^{2,3,4}$} \affiliation{$^1$Department of
Physics, Princeton University, Princeton, New Jersey, 08544}
\affiliation{$^2$Department of Chemistry, Princeton University,
Princeton, New Jersey, 08544} \affiliation{$^3$Program in Applied and
Computational Mathematics and PRISM, Princeton, New Jersey, 08544}
\affiliation{$^4$Princeton Center for Theoretical Physics, Princeton,
New Jersey, 08544} \pacs{} \date{\today}

\begin{abstract}
We have devised an isotropic interaction potential that gives rise to
negative thermal expansion (NTE) behavior in equilibrium many-particle
systems in both two and three dimensions over a wide temperature and
pressure range (including zero pressure).  An optimization procedure is used in order to
find a potential that yields a strong NTE effect.  A key feature of
the potential that gives rise to this behavior is the softened
interior of its basin of attraction.  Although such anomalous behavior
is well known in material systems with directional interactions (e.g.,
zirconium tungstate), to our knowledge this is the first time that NTE
behavior has been established to occur in single-component
many-particle systems for isotropic interactions.  Using
constant-pressure Monte Carlo simulations, we show that as the
temperature is increased, the system exhibits negative, zero and then
positive thermal expansion before melting (for both two- and three-dimensional systems).  The behavior is explicitly compared to that of
a Lennard-Jones system, which exhibits typical expansion upon heating for all
temperatures and pressures.  
\end{abstract}

\maketitle

\section{Introduction}

Control of thermal expansion properties of materials is of
technological importance due the need for structures to withstand
ambient temperature variations.  It is also of fundamental interest
because of the variety and intricacy of qualitatively different
mechanisms by which materials expand or contract upon
heating.\cite{NTE-Evans} In particular, negative thermal expansion
(NTE) behavior, a well-known but unusual phenomenon in many-particle
systems, has been observed only in multi-component materials with open
unit cell structures in which the bonding of component particles is
highly directional.  Necessarily a result of anharmonicity of the
potential energy of the system, the mechanism by which a material
contracts upon heating may be highly intricate.\cite{NTE-Pryde}

In the technological realm, materials with zero thermal expansion
(those that do not expand or contract upon heating) can aid in the
longevity of space structures, bridges and piping
systems.\cite{SigmundTorquato} It has been proposed that materials
with very large thermal expansion coefficients may function as
actuators, and those with negative thermal expansion coefficents may
be of use as thermal fasteners.\cite{SigmundTorquato}

Perhaps the most common example of a solid exhibiting NTE is that of
ice, which contracts upon melting into liquid water.\cite{Fletcher} In
its solid form, hexagonal ice also undergoes negative thermal
expansion for very low temperatures.\cite{NTE-Evans, Footnote} This
behavior is a result of the volume dependence of the transverse normal
mode (phonon) frequencies of this material, characterized by a
negative {\it Grun\"eisen parameter}.\cite{AshcroftMermin} Another
example of a material that undergoes NTE is zirconium tungstate, $Zr
W_2 O_8$, which exhibits this behavior for an extremely large
temperature range, namely $0.3K$ through $1050K$.\cite{NTE-Mary}
Again, this has been attributed to the phonon properties of the
crystal, specifically low frequency phonon modes that may propagate
without distorting the $WO_4$ tetrahedra and $ZrO_6$ octahedra that
make up the structure.\cite{NTE-Pryde} Other examples of materials
that exhibit NTE behavior are $Lu_2W_3O_{12}$,\cite{Forster} diamond
and zincblende semiconductors,\cite{Scheffler} as well as
$Sc_2(WO_4)_3$.\cite{Sc2} An essential feature of the aforementioned
systems is their relative openness (low densities) as compared to
highly-coordinated structures.

Geometries for multicomponent composite systems that give rise to
extreme NTE behavior have been designed using topology optimization
methods.\cite{SigmundTorquato}  In these cellular materials, each of
three component materials has a nonnegative thermal expansion
coefficient, but upon heating, the materials (two solid phases and one
void phase) undergo overall contraction.  An essential feature of
these structures are their local non-convex (re-entrant) ``cells''.
This is an example of an {\it inverse problem} in the sense that
optimal microstructures are found that yield a targeted macroscopic
property (in this case, NTE behavior).

In previous work,\cite{RST-PRL, RST-PRE} we investigated whether
isotropic potentials could be found that produce as ground states
low-coordinate crystal structures (e.g., the honeycomb lattice in two
dimensions and the diamond lattice) using inverse optimization
methodologies.  Here, we propose a solution to a different inverse
problem: finding an isotropic pair potential that produces a classical
many-particle system in a maximally coordinated, single-component
solid state that undergoes NTE under a wide temperature and pressure
range (including zero pressure).  This is a counterintuitive
proposition because conventional wisdom suggests that open structures
composed of particles with highly directional interactions are
necessary to observe NTE.  Nonetheless, there is no fundamental reason
for excluding this phenomenon in highly-coordinated isotropic systems.
Colloidal dispersions are a possible experimental test-bed system for
NTE behavior, as in these systems interparticle interaction potentials
may be designed by combining several interaction types (e.g.,
electrostatic repulsion, depletion, and dispersion).\cite{Russel}
Although the potential proposed in this paper is not explicitly
constructed from known colloidal interactions, it is possible that
future developments will allow for such a potential to be devised.

\begin{figure}[H]
\begin{center}
\includegraphics[width=3.2in]{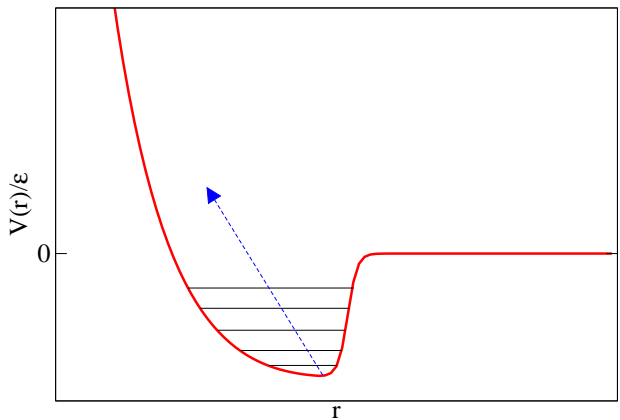}
\caption{(Color online) Schematic depiction of an isotropic pair
interaction with a softened interior in its basin of attraction.
Thermal fluctuations cause the average nearest-neighbor distance to
decrease, resulting in an overall contraction of the system upon
heating.  Here, the horizontal black lines within the basin represent
temperature values, and the blue arrow (broken line) represents the
thermally averaged nearest-neighbor distance.}\label{fig:schematic}
\end{center}
\end{figure}

We have found that a sufficient condition for a potential to give rise
to a system with NTE behavior is that it exhibits a {\it softened
interior core} within a basin of attraction (as depicted schematically
in Fig. \ref{fig:schematic}).  Such behavior of the potential may be
characterized by its third derivative with respect to position at its
minimum: if this is positive, the curvature is an increasing function
of position and the inner part of the basin is ``softer'' than the
outer part.  The potential function proposed in this paper thus
belongs to a class of functions that give rise to NTE behavior; it
will henceforth called the softened-interior-core, or ``SIC''
potential.  It is constructed by using a type of optimization
procedure wherein parameters are chosen such that the potential will
give rise to strong NTE behavior (via a large, positive third
derivative of the pair potential with respect to $r$, the distance
between particles).  As will be explained in the next section, as the
temperature is increased in such systems, the nearest-neighbor
distances fluctuate about smaller average distances, causing NTE
behavior.  To demonstrate this phenomenon, we perform Gibbs-ensemble
(constant number of particles, temperature and pressure, or ``NPT'')
Monte Carlo \cite{Smit} simulations in both two and three dimensions
in which the volume of the system is computed as a function of both
ambient temperature and pressure.  We find that at low temperature,
the system has its strongest negative thermal expansion.  As the
temperature is increased, the system contracts further until it
reaches a minimum in volume (or area), and then expands as a function
of temperature before melting.  As a basis for comparison, we perform
similar calculations using the Lennard-Jones potential, which exhibits
conventional behavior in that it only expands as temperature is
increased.
\vspace{0.5in}
In Section II, we discuss how the SIC potential was constructed.  In
Section III, we discuss the details and results of the energetics
calculations and NPT simulations, for both the SIC potential and for
the Lennard-Jones potential.  In Section IV, we summarize the results
and discuss future work pertaining to inverse problems in
many-particle systems.

\section{Theoretical Background}

We consider classical $N$-particle systems (in
two and three dimensions) with interactions that are
pairwise additive and isotropic.  Denoting the positions of the
particles by $\rv_1,...,\rv_N$ and their momenta by $\pv_1,...,\pv_N$,
the Hamiltonian of the system is
\begin{equation}
H(\rv_1,...,\rv_N, \pv_1,...,\pv_N) = \sum_{i<j}^{N}V_I(|\rv_i-\rv_j|) +
\sum_{i=1}^N \frac{\pv_i^2}{2m},
\end{equation} 
where $V_I$ is a interaction potential that is a purely radial function, and $m$ is the mass of each
particle, taken as unity for the remainder of this paper.  Henceforth we employ the reduced temperature of the system, $T^* = k_B
T/\epsilon$, where $T$ is the temperature, $k_B$ is Boltzmann's
constant, and $\epsilon$ is a parameter in the pair potential with
units of energy.

The isobaric thermal expansion coefficient $\alpha$ is defined as
\begin{equation}
\alpha = \frac{1}{V}\left(\frac{\partial V}{\partial T}
\right)_p,\label{thermal-expansion-coefficient}
\end{equation}
where $V$ is the volume of the system, and $p$ is pressure.  Of course, $\alpha$ is negative if the
system contracts upon heating.  We seek an interaction potential that
gives rise to this effect in an equilibrium sense.

\subsection{Construction of the SIC potential}
\begin{figure}[H]
\begin{center}
\includegraphics[width=4in]{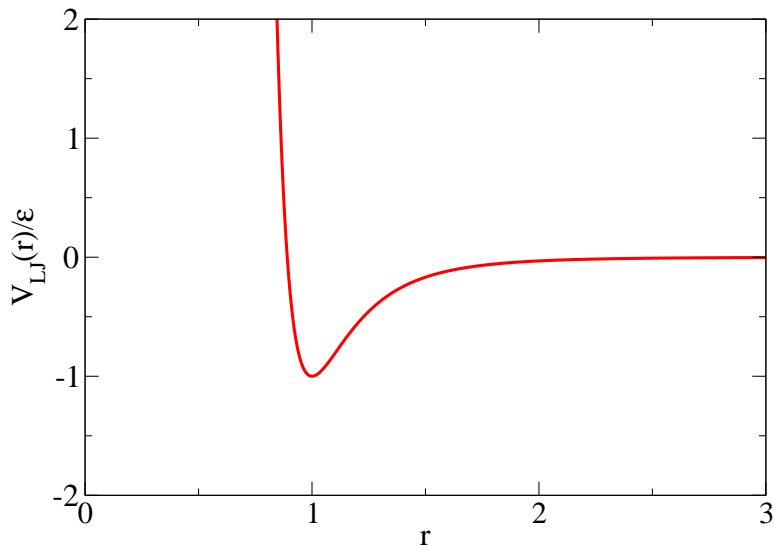}
\caption{(Color online) The Lennard-Jones potential, as given in
Eq. (\ref{lj}), with $\epsilon_{LJ}=\epsilon$ and $\sigma_{LJ} = 1/2^{1/6}$.  As
can be easily seen, the curvature of the potential is greater to the
left of the minimum at $r=1$ than to the right.  As a result, when the
temperature of a system interacting via this potential is increased,
thermal fluctuations cause the average nearest-neighbor distance to
increase.}\label{fig:potential-LJ}
\end{center}
\end{figure}

\begin{figure}
\begin{center}
\includegraphics[width=4in]{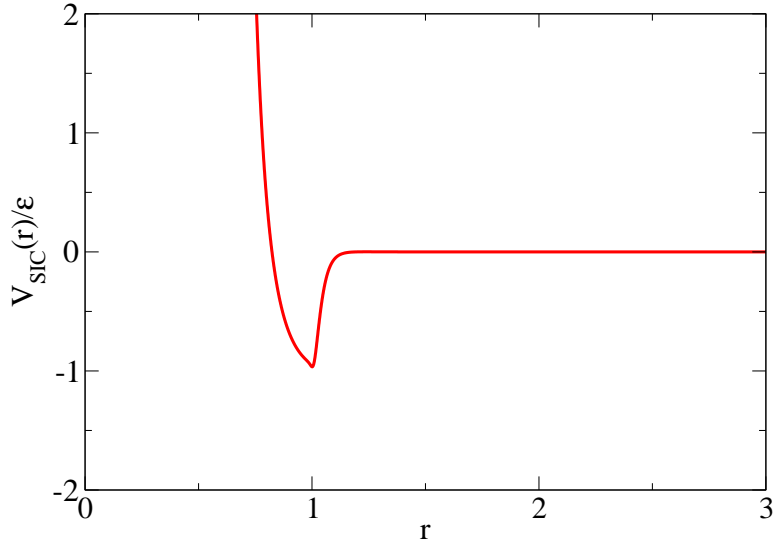}
\caption{(Color online) The SIC potential, as given in
Eq. (\ref{mm}), with the parameters $A=9.30865$, $B=0.1$, $C=9A$,
$r_M=1$, $\epsilon_M=\epsilon$ and $\sigma_M=0.5$.  As can easily be seen, the curvature of
the potential is greater to the right of the minimum at $r=1.00069$ than
to the left.  As a result, when the temperature of a system interacting via this
potential is increased, thermal fluctuations cause the average
nearest-neighbor distance to decrease.}\label{fig:potential-concocted}
\end{center}
\end{figure}

In order to motivate how we devise a pair interaction potential that
gives rise to negative thermal expansion for two- and
three-dimensional many-particle systems in thermal equilibrium,
consider first the Lennard-Jones potential:
\begin{equation}
V_{LJ}(r) = 4\epsilon_{LJ} \left[ \left(\frac{\sigma_{LJ}}{r}\right)^{12} -
\left(\frac{\sigma_{LJ}}{r}\right)^6\right],\label{lj}
\end{equation}    
where $\epsilon_{LJ}>0$ and $\sigma_{LJ}>0$ are energy and length
parameters, respectively.  This potential is shown in
Fig. \ref{fig:potential-LJ}, with $\epsilon_{LJ}=\epsilon$ and
$\sigma_{LJ} = 1/2^{1/6}$.  We adopt this parameter choice for the
Lennard-Jones potential for the remainder of this paper.  For a system
of $N$ particles interacting via the Lennard-Jones potential, where
both the temperature and pressure are zero, the equilibrium
configuration in two dimensions is the triangular lattice, and in
three dimensions it is a maximally coordinated lattice (the hexagonal
close-packed, or hcp) with the nearest neighbors lying within the
basin of attraction of the potential.  Clearly, the curvature of the
potential to the left of the minimum at $r=1$ is significantly greater
than that to the right.  This is reflected in the negative third
derivative of the potential at its minimum, i.e., $V_{LJ}^{(3)}(1)=
-1512\epsilon$.  As a result, as the temperature is increased from
zero, thermal fluctuations cause the distribution of nearest-neighbor
distances to be skewed to the right.  Thus, in thermal equilibrium,
the particles are farther apart on average, and so the system should
expand upon a temperature increase, i.e., $\alpha>0$.  The behavior of
the system should not qualitatively change if the pressure is
positive.  This is the mechanism for positive thermal expansion in
this system.
\vspace{0.5in}
Here, we construct an isotropic pair potential with a softened
interior by finding a function with a strongly positive third
derivative (i.e., increasing curvature) within a basin of attraction
(which should result in NTE behavior, as described in Section I).  We
choose the Morse potential as a starting point, which is given by
\begin{equation}
V_M(r) =
\epsilon_M\left\{\exp\left[2(r_M-r)/\sigma_M\right]-2\exp\left[
(r_M-r)/\sigma_M\right]\right\},\label{morse}
\end{equation}
where $r_M>0$ and $\sigma_M>0$ are adjustable parameters.  It has a
basin of attraction with a minimum at $r_M$, and its third derivative
evaluated at its minimum is negative, i.e.,
$V^{(3)}_M(r_M)=-6\epsilon_M/\sigma_M^3$.  Since we seek a potential with a
positive third derivative in its basin, modification of this
potential is necessary.  We thus use a rescaling function for distance defined by
\begin{equation}
r^*(r)=Ar+B\left\{ \log\cosh \left[C(r-r_M)\right] -\log\cosh (C r_M)\right\},\label{Scaling}
\end{equation}
where $A>0$, $B>0$, and $C>0$ are free parameters such that $A-BC>0$.
This function has the property that for $r<r_M$, it asymptotes to a
straight line going through the origin with slope $A-BC$, but for
$r>r_M$, it quickly asymptotes to a straight line of slope $A+BC$.  We
define the softened-interior-core (SIC) potential to be
\begin{equation}
V_{SIC}(r) = \epsilon_M\left(\frac{0.8}{r}\right)^{15} + V_M (r^*(r)).\label{mm}
\end{equation}
This potential is plotted in Fig. \ref{fig:potential-concocted} with
$A=9.30865$, $B=0.1$, $C=9A$, $r_M=1$, $\epsilon_M=\epsilon$ and
$\sigma_M=0.5$.  With this choice of parameters, the function is such
that its curvature is greater to the right of its minimum, at
$r=1.00069$, than to the left, and it has a third derivative of
$V^{(3)}_{SIC}(r_M)=1.13660\times 10^5\epsilon_M$.  Note that the
parameter values have been chosen such that this third derivative is
quite large, with the intention of producing strong NTE behavior.  In
this sense, optimization of the potential is performed over the space
of functions defined by Eqs. (\ref{Scaling}) and (\ref{mm}).  The
first term on the right side in Eq. (\ref{mm}) yields a stiffer core
than would otherwise have been present.  As we show in Section III.A.,
this has the effect of causing a maximally coordinated structure
(face-centered cubic, or fcc) to be the energetically stable structure
at zero pressure.  This ensures that the nearest neighbor is located
near the bottom of the potential well.  Farther neighbors interact
only very weakly.  Due to the softened interior within the basin of
attraction of the SIC potential, given in Eq. (\ref{mm}),
thermal fluctuations cause nearest neighbor distances to decrease, on
average.  This has the effect of causing an overall contraction of the
system.  Note that this argument is independent of whether the system
in question is two or three dimensional.  Thus, we expect the same
potential (\ref{mm}) to yield NTE behavior in both dimensions.

This rescaling procedure could have been applied to the Lennard-Jones
potential rather than the Morse potential.  However, it was found that
since the former possesses a much more strongly negative third
derivative at its minimum than the latter (when their minima are at
the same position and depth), the Morse potential is more suitable to
exhibiting a strongly increasing curvature within its attractive basin.
\vspace{0.5in}
\section{Results and Discussion}
In this section we present lattice sums (Madeling energies as a
function of specific area or volume for a number of crystal
structures), for the Lennard-Jones and SIC potential
systems in two and three dimensions.  We thus determine the low
temperature thermodynamically stable crystal structure for each and
using the Maxwell construction,\cite{LandauLifshitzStatMech} we
ascertain the zero-temperature range of stability in pressure.  In
this section we take $\epsilon=1$.
\subsection{Energetics of the Lennard-Jones and SIC potential}

\begin{figure}
\begin{center}
\includegraphics[width=4in]{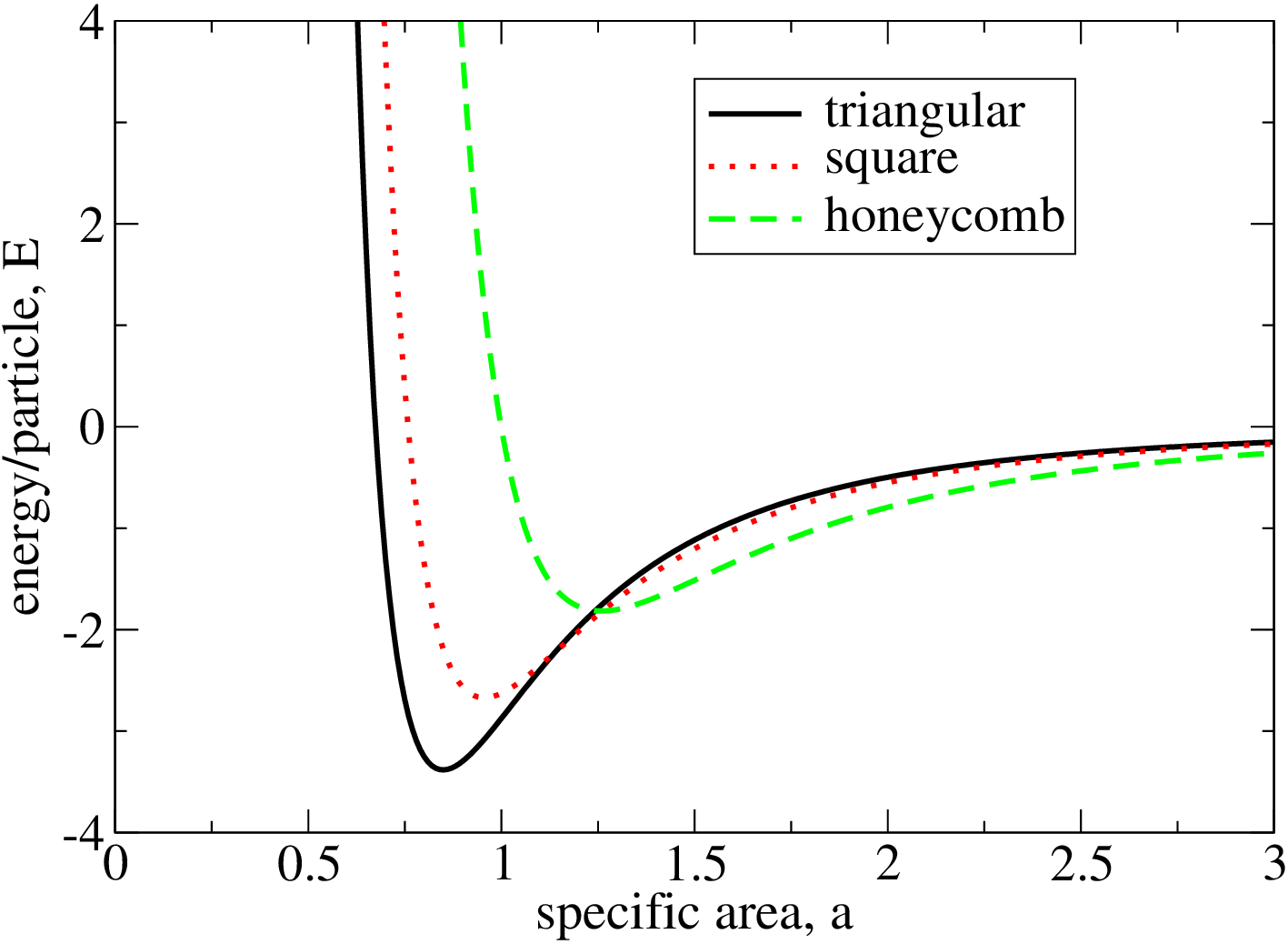}
\caption{(Color online) Two-dimensional lattice sums for the Lennard-Jones potential
including the triangular, square and honeycomb lattices.  The lowest
overall energy stucture is the triangular lattice, which is thus the
stable structure at zero pressure.  From the Maxwell construction, we
find that the pressure range of stability at zero temperature is
$0<p<\infty$, and the range of stability in specific area is
$0<a<0.85$.  }\label{fig:latticesums-LJ-2D}
\end{center}
\end{figure}

\begin{figure}
\begin{center}
\includegraphics[width=4in]{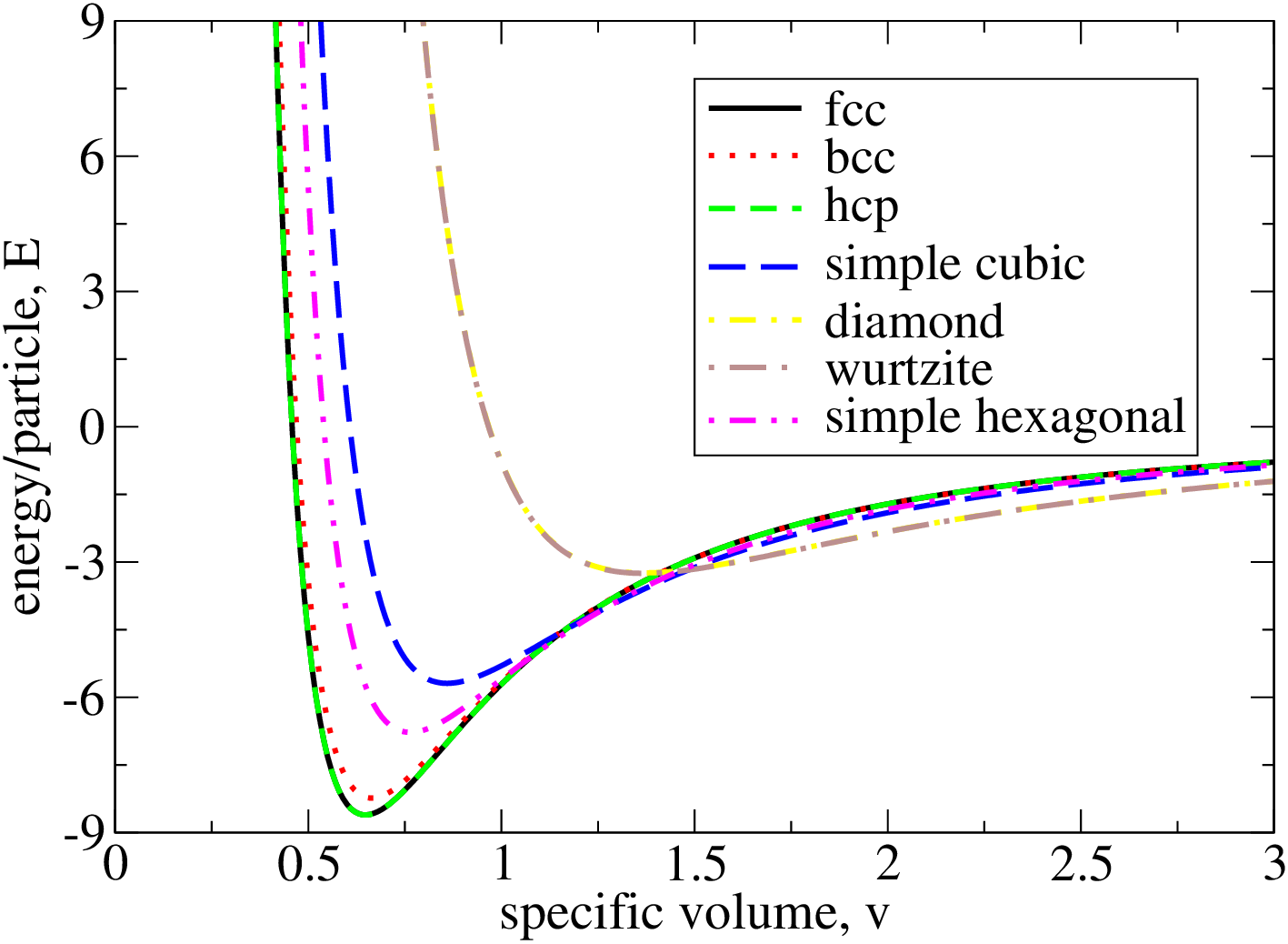}
\caption{(Color online) Three-dimensional lattice sums for the Lennard-Jones
potential including the fcc, bcc, hcp, simple cubic, diamond,
wurtzite, and simple hexagonal lattices.  The lowest overall energy
stucture is the hcp lattice, which is thus the stable structure at
zero pressure.  From the Maxwell construction, we find that the
pressure range of stability at zero temperature is $0<p<1240$, and
the range of stability in specific area is
$0.33<v<0.65$. }\label{fig:latticesums-LJ-3D}
\end{center}
\end{figure}

\begin{figure}
\begin{center}
\includegraphics[width=4in]{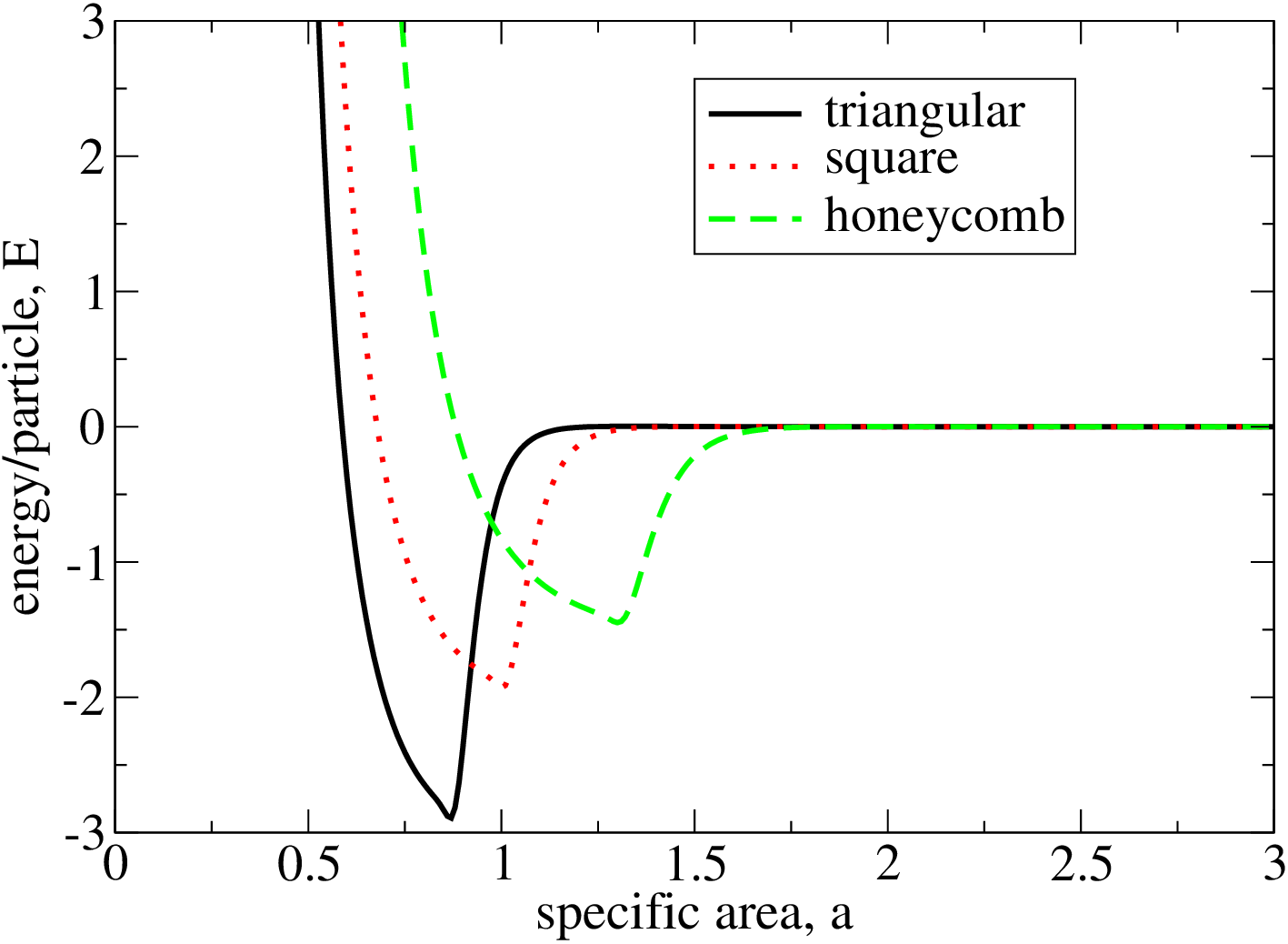}
\caption{(Color online) Two-dimensional lattice sums for the SIC potential including the triangular, square and honeycomb
lattices.  The lowest overall energy structure is the triangular
lattice, which is thus the stable structure at zero pressure.  From
the Maxwell construction, we find that the pressure range of stability
at zero temperature is $0<p<\infty$, and the range of stability in
specific area is $0<a<0.87$. }\label{fig:latticesums-concocted-2D}
\end{center}
\end{figure}

\begin{figure}
\begin{center}
\includegraphics[width=4in]{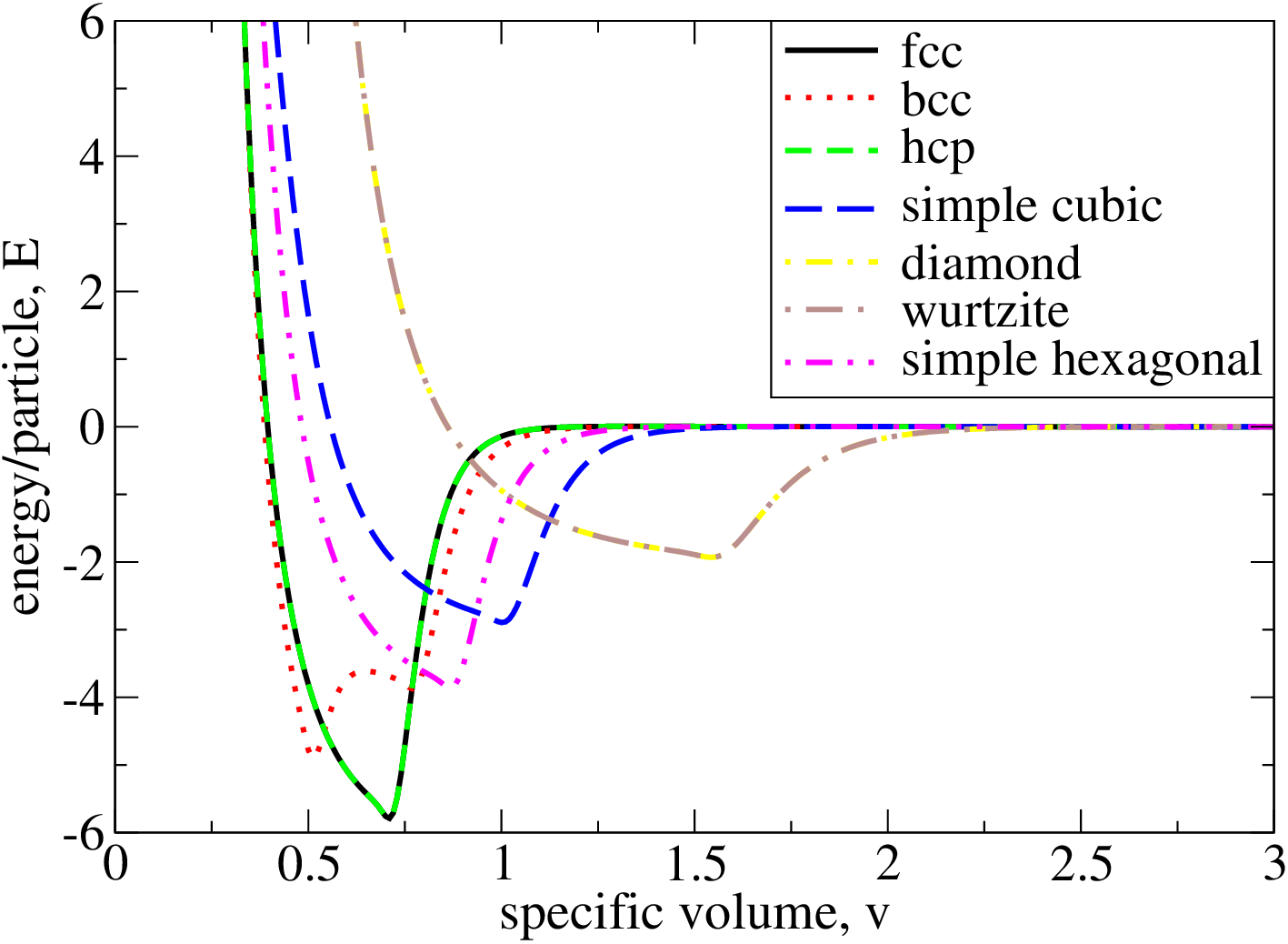}
\caption{(Color online) Three-dimensional lattice sums for the SIC
potential including the fcc, bcc, hcp, simple cubic, diamond,
wurtzite, and simple hexagonal lattices.  The lowest overall energy
stucture is the hcp lattice, which is thus the stable structure at
zero pressure.  From the Maxwell construction, we find that the
pressure range of stability at zero temperature is $0<p<4.5$, and the
range of stability in specific area is $0.70<v<0.71$.  There is
another region of stability of this system at significantly smaller
specific volume, but this is irrelevant to the present
work.}\label{fig:latticesums-concocted-3D}
\end{center}
\end{figure}

The lattice sums for the Lennard-Jones systems in two and three
dimensions are shown in Figs. \ref{fig:latticesums-LJ-2D} and
\ref{fig:latticesums-LJ-3D}, respectively.  In the two-dimensional
case, the Madelung energies of three crystal structures (the
triangular lattice, square lattice, and honeycomb lattice) are plotted
as a function of specific area, $a$.  The lowest overall energy
stucture is the triangular lattice, which is thus the most stable
structure at zero pressure.  From the Maxwell construction, we find
that the pressure range of stability at zero temperature is
$0<p<\infty$, and the range of stability in specific area is
$0<a<0.85$.  The lattice sums for the three-dimensional system are
plotted in Fig. \ref{fig:latticesums-LJ-3D}, with the Madeling
energies of seven crystals being plotted (fcc, body-centered
cubic--bcc, hcp, simple cubic, diamond, wurtzite, and simple hexagonal
lattices).  The maximally coordinated structures, namely fcc and hcp,
clearly have the lowest energies overall, and are thus the stable
structures at zero pressure.  These two lattices are extremely close
in energy, although the hcp has the lower energy for specific areas
for which the nearest neighbor is near the bottom of the basin of
attraction, which is the regime we study in this paper.  Its pressure
range of stability at zero temperature is $0<p<1240$, and its range of
stability in specific volume is $0.33<v<0.65$.
\vspace{0.5in}
\vspace{0.5in}
Lattice sums for the SIC potential systems in two and three dimensions
(with parameters $A=9.30865$, $B=0.1$, $C=9A$, $r_M=1$,
$\epsilon_M=\epsilon$ and $\sigma_M=0.5$) are shown in
Figs. \ref{fig:latticesums-concocted-2D} and
\ref{fig:latticesums-concocted-3D}.  For both cases, the same lattices
are plotted as in the Lennard-Jones case.  As in the two-dimensional
Lennard-Jones system, the triangular lattice has the lowest overall
energy and is thus the zero-pressure stable structure.  The pressure
range of stability at zero temperature is $0<p<\infty$ and the range
of stability in specific area is $0<a<0.87$.  In the three-dimensional
SIC system, the maximally coordinated structures are again the most
stable, and extremely close in energy.  However, in this case it is
the fcc lattice which has the lowest overall energy.  Note that the
stiff-core term in the potential $V_{SIC}$, the first term on the
right-hand side of Eq. (\ref{mm}) is essential for the maximally
coordinated structures to have lower energies than the other lattices.
Without this term, lower-coordinated structures are energetically
stable at specific volumes for which several neighbor distances may
fall within the basin of attraction of the potential.  It follows that
the argument made in Section II.A. motivating construction of the SIC
potential as a candidate for NTE behavior would not hold in this case.

In three dimensions, there are two distinct pressure and density
regions in which the fcc lattice is stable, however, we are only
interested in the one in which the nearest neighbor falls within the
basin of attraction of the potential.  For this region, the pressure
range of stability at zero temperature of the fcc lattice in the
three-dimensional case is $0<p<4.5$, and the specific volume range of
stability is $0.70<v<0.71$.

\vspace{0.5in}

\vspace{0.5in}

\subsection{NPT Monte Carlo simulation results}

Monte Carlo simulations were run in the NPT ensemble on the two- and
three-dimensional Lennard-Jones and SIC systems for a
number of pressures and temperatures.  In all simulations, Monte Carlo
sampling is carried out until equilibrium is achieved, after which the
fluctuating area/volume is repeatedly sampled until a sufficiently
precise average is obtained.  In both two-dimensional systems, the
triangular lattice was used with $N=1020$ particles in a rectangular
simulation cell with periodic boundary conditions imposed.  In both
three-dimensional systems, the fcc lattice was used with $N=864$ in a
cubic cell also with periodic boundary conditions.  Note that in
the case of the Lennard-Jones system, the hcp, not the fcc, lattice is
the thermodynamically stable structure.  However, we employ the latter
in simulations in order to directly compare the results to those of
the SIC system.  This choice is justified due to the
extreme closeness in Madelung energies of the fcc and hcp lattices.
To confirm that the maximally coordinated lattices were indeed the ground
states for each of the four systems discussed here, annealing
simulations were performed in which each system was cooled through its
freezing point.  In each case, a maximally coordinated lattice resulted as the
appropriate equilibrium crystal state.  

\begin{figure}
\begin{center}
\includegraphics[width=4in]{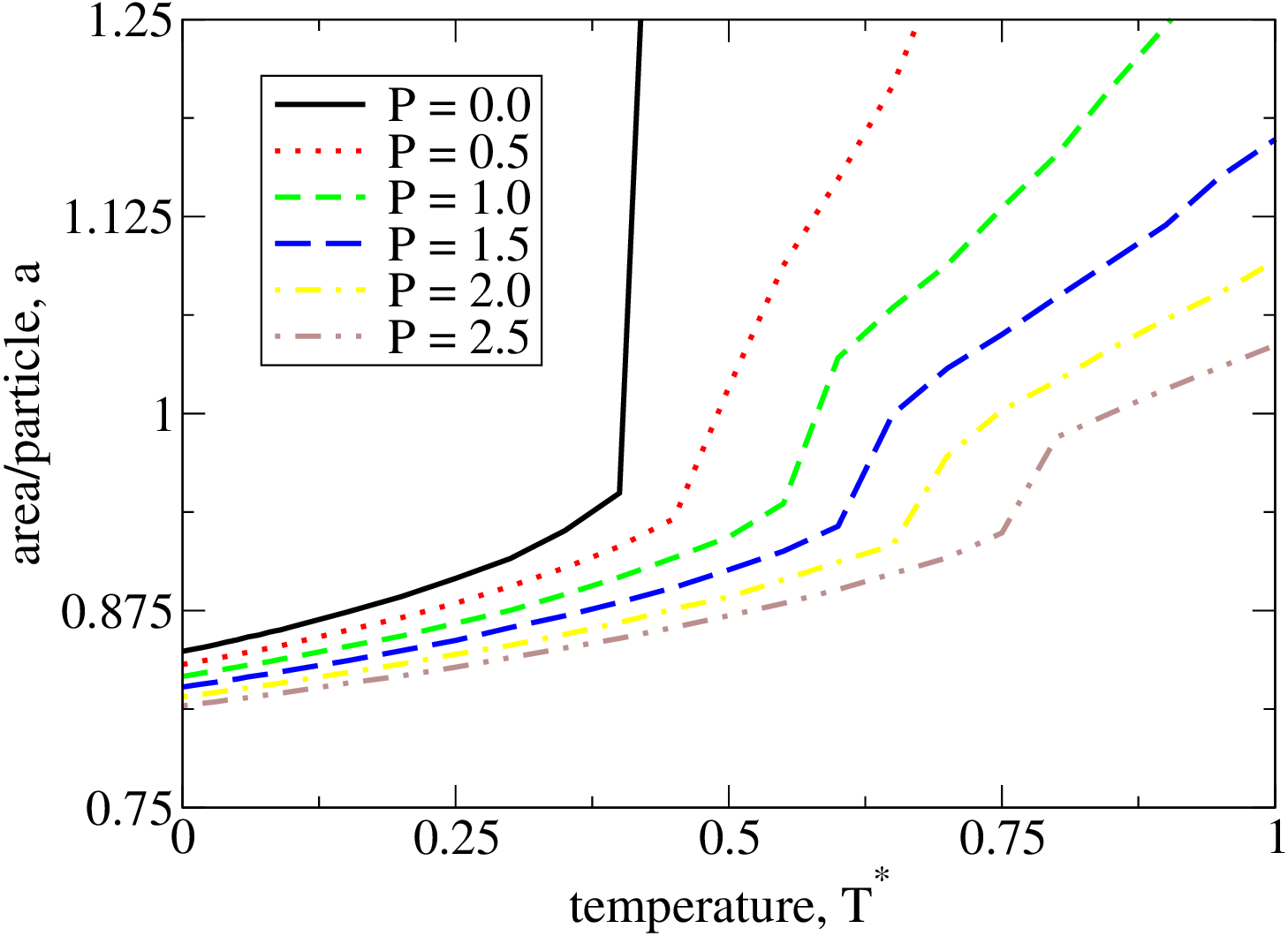}
\caption{(Color online) Specific area as a function of temperature for
a number of different pressure values for the two-dimensional
Lennard-Jones system.  These results are obtained from NPT Monte Carlo
simulations with $N=1040$ particles and periodic boundary conditions.
In its crystalline (triangular) state, this system expands when temperature is
increased for all pressures and temperatures.  Note that kinks
in the curves indicate melting of the crystal.  }\label{fig:VT-LJ-2D}
\end{center}
\end{figure}

\begin{figure}
\begin{center}
\includegraphics[width=4in]{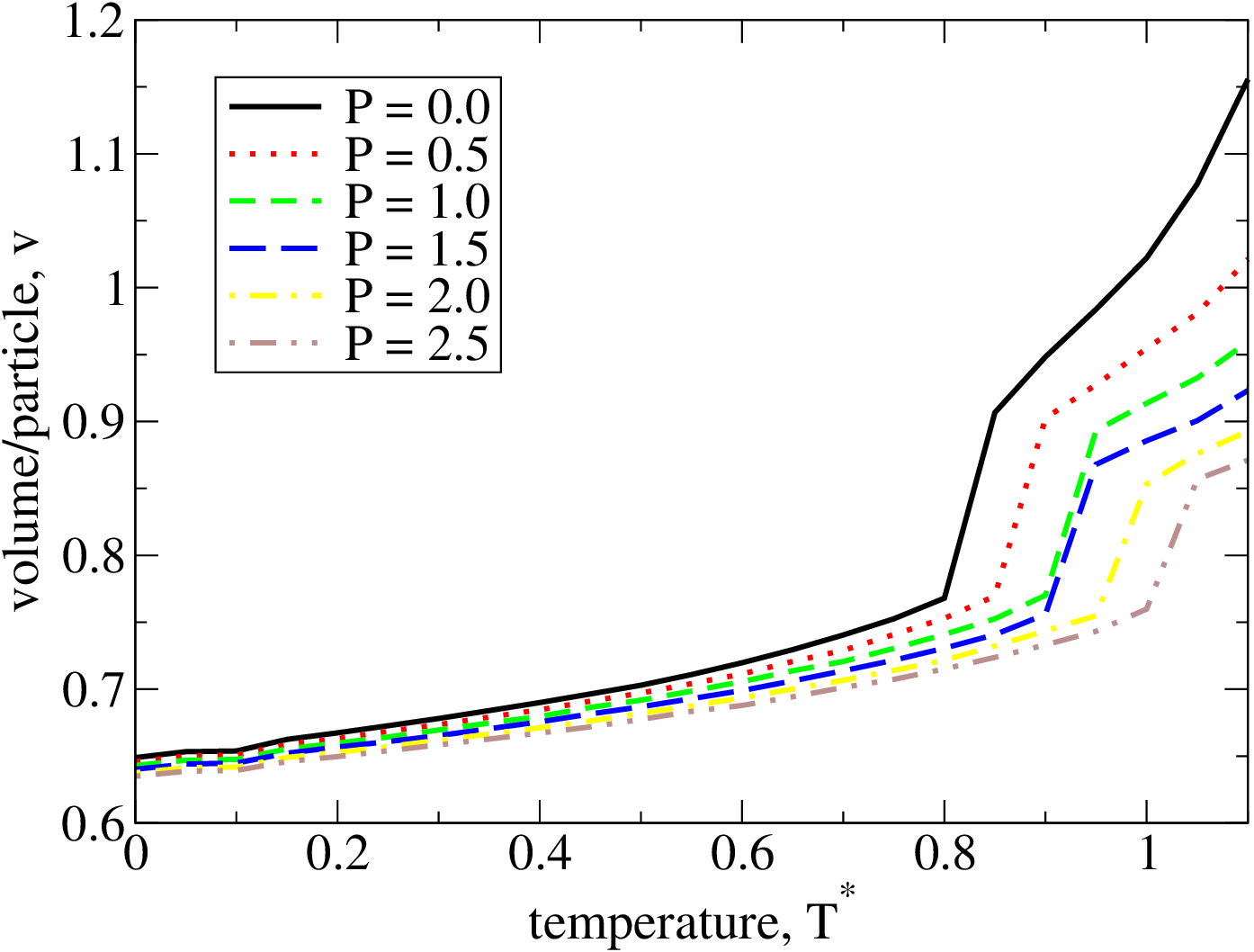}
\caption{(Color online) Specific volume as a function of temperature for
a number of different pressure values for the three-dimensional
Lennard-Jones system.  These results are obtained from NPT Monte Carlo
simulations with $N=864$ particles and periodic boundary conditions.
In its crystalline (fcc) state, this system expands when temperature is
increased for all pressures and temperatures.  Note that the ``kinks''
in the curves indicate melting of the crystal.  }\label{fig:VT-LJ-3D}
\end{center}
\end{figure}

The area/volume dependence on temperature of the Lennard-Jones systems
in two and three dimensions is shown in Figs. \ref{fig:VT-LJ-2D} and
\ref{fig:VT-LJ-3D}, respectively.  The plots show results for a number
of pressure values, $p=0,0.5,1.0,1.5,2.0$ and $2.5$.  The data show
that in both cases, the solid expands until it melts.  The sharp bends, or ``kinks''
in the curves indicate the first-order melting transition.

\vspace{0.5in}

\vspace{0.5in}

\begin{figure}
\begin{center}
\includegraphics[width=4in]{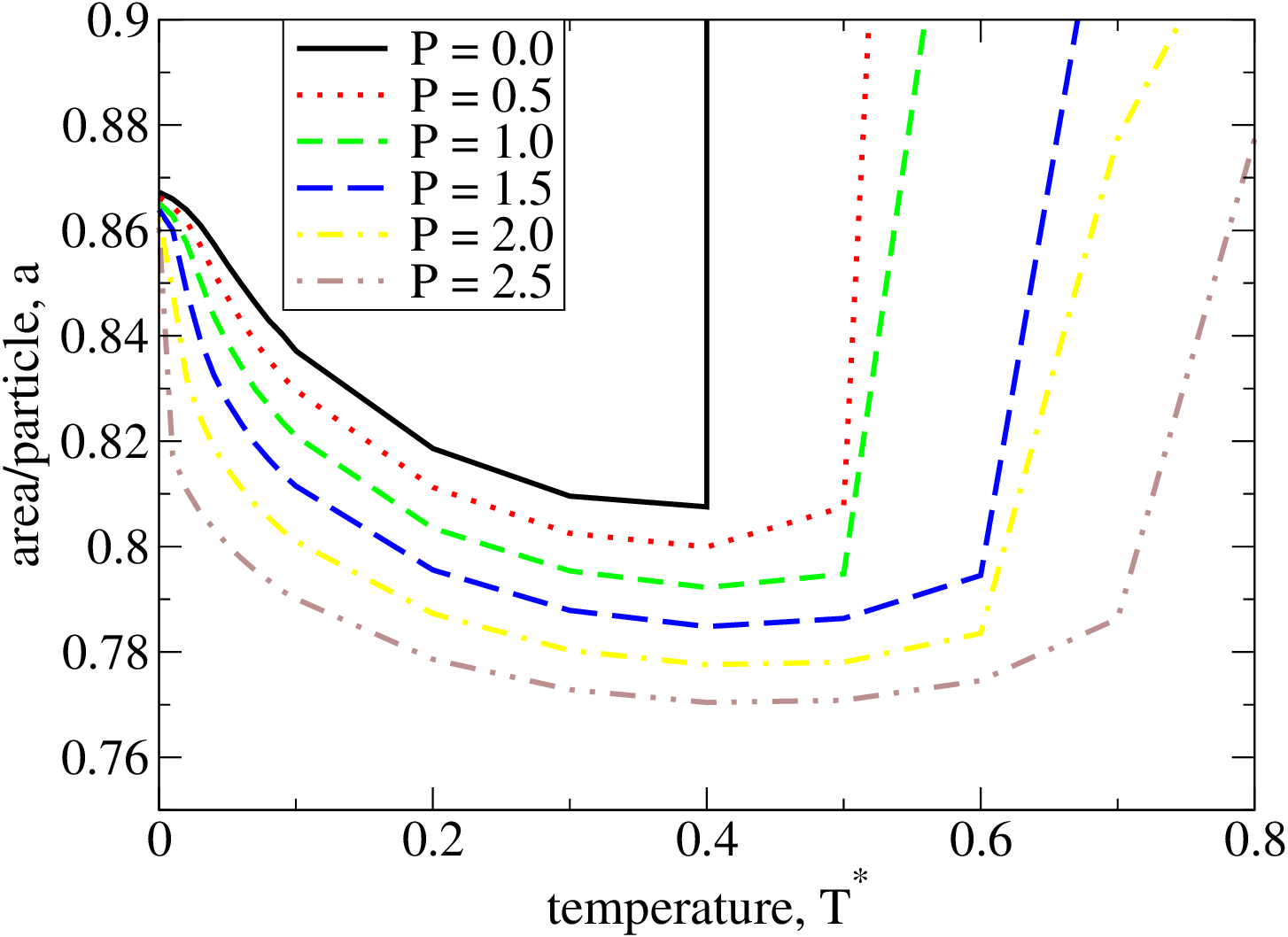}
\caption{(Color online) Specific area as a function of temperature for
a number of different pressure values for the two-dimensional SIC system.  These results are obtained from NPT Monte Carlo
simulations with $N=1040$ particles and periodic boundary conditions.
In its crystalline (triangular) state, this system undergoes NTE for
low temperatures, but as temperature is increased, thermal expansion
passes through zero and then becomes positive.  Note that the
``kinks'' in the curves indicate melting of the crystal.
}\label{fig:VT-concocted-2D}
\end{center}
\end{figure}

\begin{figure}
\begin{center}
\includegraphics[width=4in]{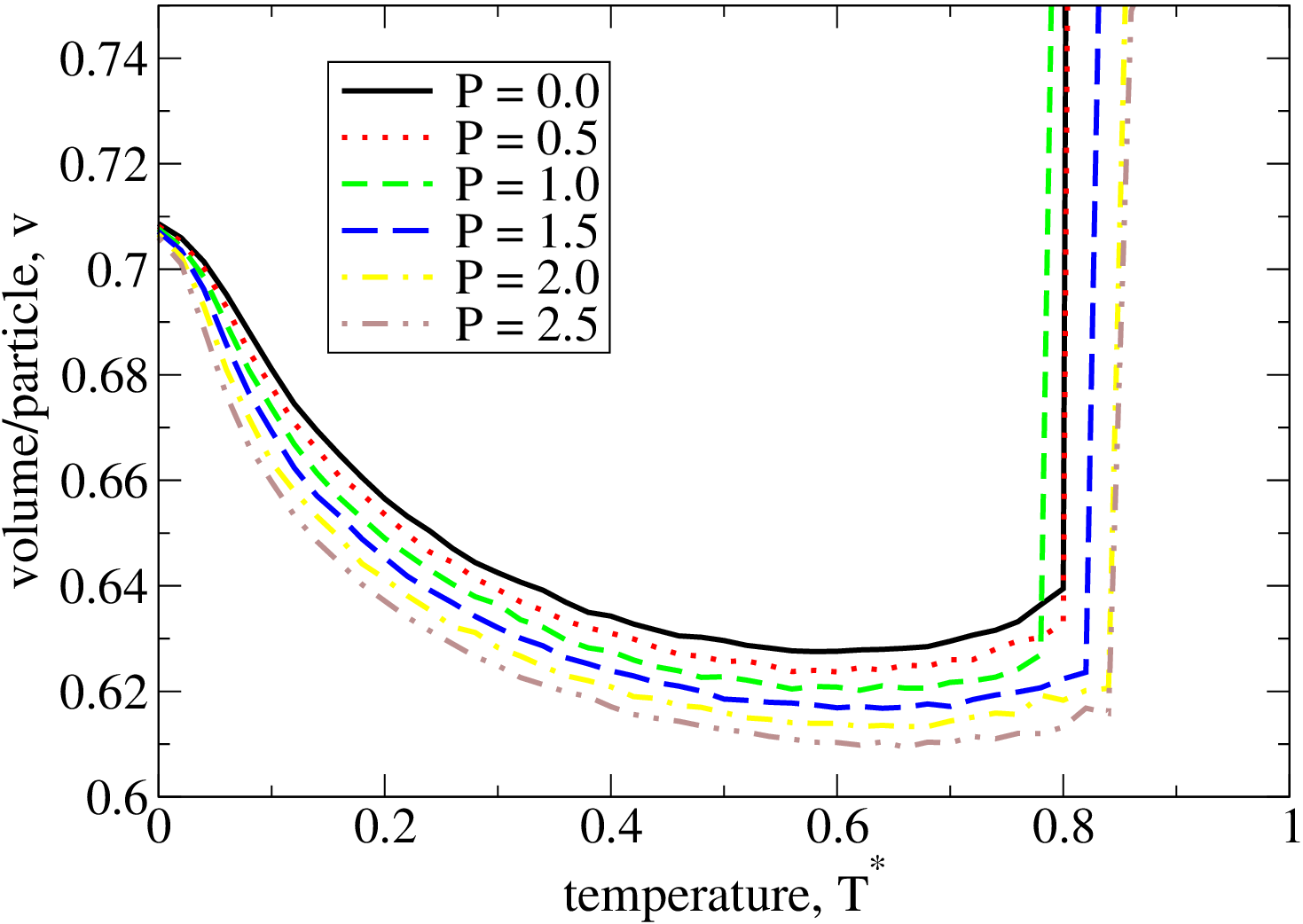}
\caption{(Color online) Specific volume as a function of temperature for
a number of different pressure values for the three-dimensional SIC system.  These results are obtained from NPT Monte Carlo
simulations with $N=864$ particles and periodic boundary conditions.
In its crystalline (fcc) state, this system undergoes NTE for
low temperatures, but as temperature is increased, thermal expansion
passes through zero and then becomes positive.  Note that the
``kinks'' in the curves indicate melting of the crystal.  }\label{fig:VT-concocted-3D}
\end{center}
\end{figure}

Simulation results for the SIC potential systems in two and
three dimensions (with parameter values $A=9.30865$, $B=0.1$, $C=9A$,
$r_M=1$, $\epsilon_M=\epsilon$ and $\sigma_M=0.5$) are shown in
Figs. \ref{fig:VT-concocted-2D} and \ref{fig:VT-concocted-3D}
respectively.  The data show that for low temperature and for each
pressure value (among $p=0,0.5,1.0,1.5,2.0$ and $2.5$), in both two
and three dimensions, the system contracts.  At a certain
pressure-dependent temperature, the thermal expansion coefficent
passes through zero and becomes positive (at the minima of the
curves).  At higher temperatures, the system undergoes positive
thermal expansion until it melts.  The temperature at which zero
thermal expansion is achieved is an increasing function of the
pressure.

\vspace{0.5in}

\vspace{0.5in}

\begin{figure}
\begin{center}
\includegraphics[width=4in]{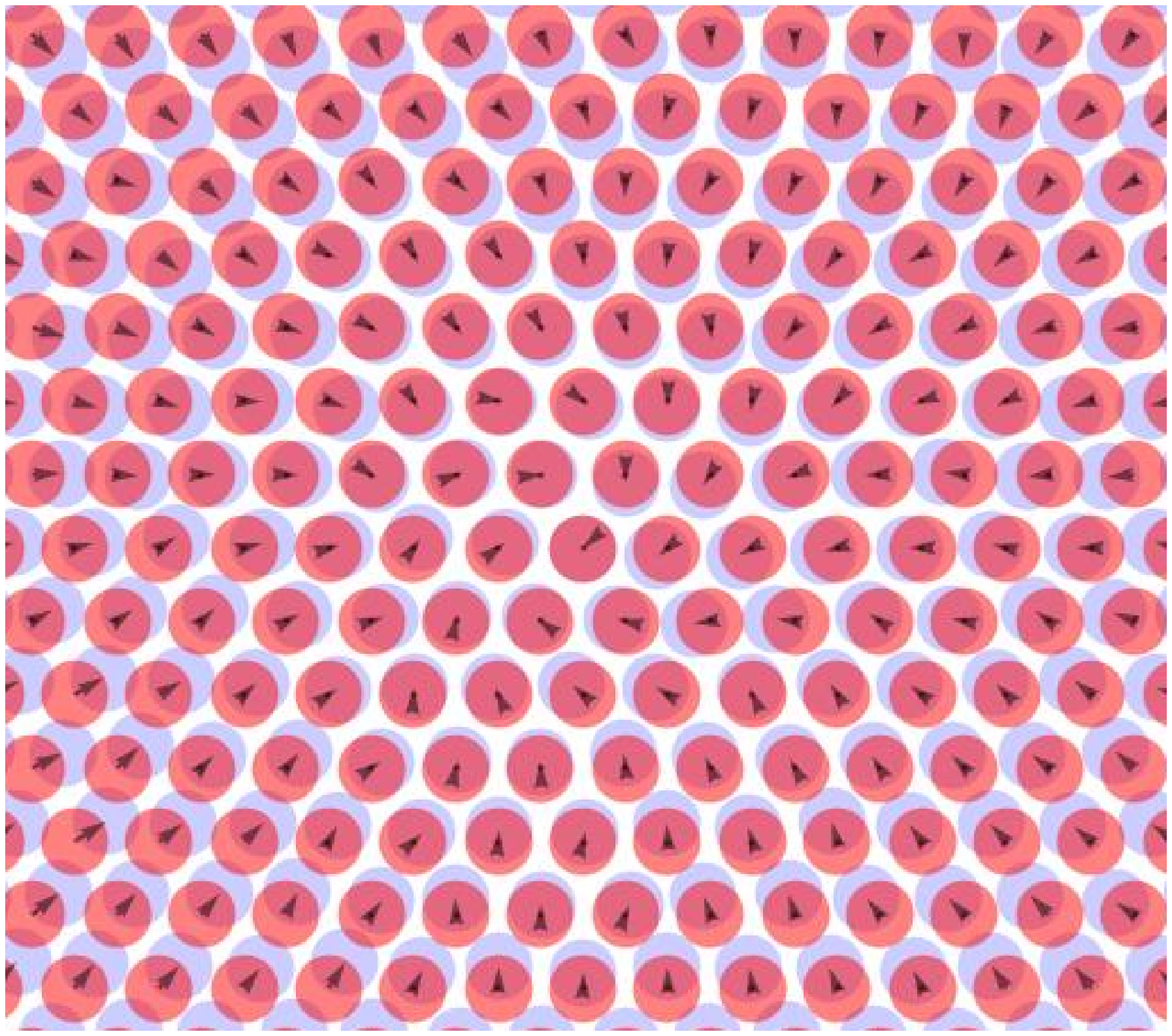}
\caption{(Color online) A section of two configurations of the
SIC system, one at $\Tstar=0.0$ (red) and the other at $\Tstar=0.3$
(blue).  The configurations are snapshots taken from NPT Monte Carlo
simulations.  Arrows indicate the displacement of the particles upon
heating from the former temperature to the latter.  The ambient
pressure is $p=1.0$.  The system appears to undergo a simple
rescaling. }\label{fig:config-concocted}
\end{center}
\end{figure}

For the two-dimensional SIC system, we show in
Fig. \ref{fig:config-concocted} two configurations of the system at
temperatures $\Tstar=0.0$ and $\Tstar=0.3$, with both at pressure $p=1.0$.  Arrows
show the direction that the particles move when the temperature is
increased from $\Tstar=0.0$ to $\Tstar=0.3$.  

\vspace{0.5in}

\begin{figure}[H]
\begin{center}
\includegraphics[width=4in]{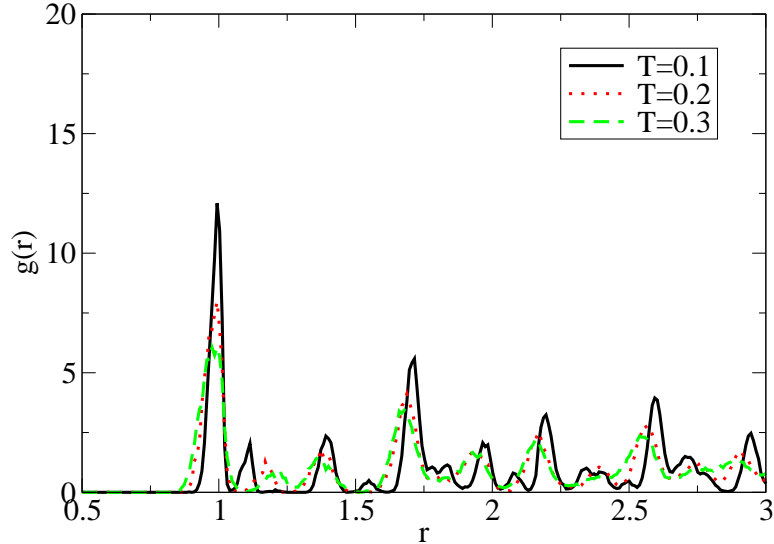}
\caption{(Color online) Radial distribution function for the SIC system at $p=1.0$ for three temperatures, $\Tstar=0.1, 0.2$ and
$0.3$.  As a result of the increasing curvature of the SIC
potential, given in Eq. (\ref{mm}), near its minimum, the first peak
is increasingly skewed to smaller $r$ as the temperature is
increased. The second peak is not present in the perfect fcc lattice, and
represents a shearing of the system.}\label{fig:rdf-concocted}
\end{center}
\end{figure}

The radial distribution function, $g(r)$, is plotted for the
three-dimensional case in Fig. \ref{fig:rdf-concocted}, for pressure
$p=1.0$ and temperatures $\Tstar=0.1,0.2$ and $0.3$.  While the first peak
(near $r=1$) appears to reach its maximum at nearly the same distance
for each temperature, the left tail of the first peak extends further
and further to the left as temperature is increased.  This is an
indicator of the NTE mechanism of the system.  Here, thermal
fluctuations cause the nearest-neighbor distance distribution to
widen, but the increasing curvature as $r$ increases of the potential, $V_{SIC}$, causes
this distribution to be skewed towards lower $r$.  The average volume
of the system decreases as a result.  Note that this is exactly the
mechanism of negative thermal expansion in the two-dimensional
SIC system as well.  Figure \ref{fig:rdf-concocted} also
shows an anomalous second peak not present in the radial distribution
function of the fcc lattice.  This indicates that in additon to the
volume change, the system undergoes a shearing as temperature is
increased from zero.

\vspace{0.5in}

\section{Conclusions}

In this paper, we have proposed an isotropic pair interaction
potential for a system of classically interacting particles that gives
rise to negative thermal expansion in maximally coordinated lattices
in both two and three dimensions.  Previously, this phenomenon had
only been observed in low-density open crystals \cite{NTE-Evans} and
in three-component composites.\cite{SigmundTorquato}  The key feature
of the proposed potential, $V_{SIC}(r)$, which is based on a
modification of the Morse potential, is a basin of attraction wherein
the curvature is an increasing function of $r$.  With this property,
thermal fluctuations cause the nearest-neighbor distances in the
maximally-coordinated lattices to decrease on average, resulting in an
overall contraction.

Using NPT Monte Carlo simulations, we compared the behavior of systems
interacting via $V_{SIC}(r)$ to those interacting via the standard
Lennard-Jones interaction potential.  We found that NTE remained
present for the former over a large temperature and pressure range,
with zero pressure included in the latter.  At sufficiently high
temperatures (still below the melting point), the thermal expansion
coefficent goes to zero and then becomes positive.  The proposed
potential may be seen as a possible solution to an inverse problem:
that of finding a microscopic interaction that yields a desired
macroscopic property.  Indeed, by adjusting the parameters of the
Morse potential, given in Eq. (\ref{morse}) and in the rescaling
function given in Eq. (\ref{Scaling}), the thermal expansion
coefficient may be manipulated.

In future work, we aim to continue the inverse problem program of
finding isotropic interaction potentials that yield systems with
targeted material properties.  One example of such a problem is
finding isotropic potentials that favor disordered systems as well as
certain types of defects in crystalline solids.  Another example is
the problem of designing an isotropic interaction potential that
yields a material with negative Poisson ratio, i.e., wherein
compression in one direction causes compression in the orthogonal direction.  This
effect has been studied experimentally and theoretically in composite
materials,\cite{Whitesides} in foam structures,\cite{FoamsTheory,FoamsExperiment} and in atomic solids.\cite{Silica}  Another challenging open problem is to find an
isotropic pair potential that produces a substance that contracts upon
melting and then continues to contract over some temperature range (as
is the case in water).  Lastly, there is the challenge of designing an
isotropic pair potential that yields a system that freezes into a
crystalline state upon an increase in temperature (``inverse
melting''), over a wide temperature and pressure range.  Both isotopes
of helium exhibit this property,\cite{Helium,Helium2} and simulations
have shown that a modified Gaussian-core interaction does as well.\cite{StillingerDebenedetti}  This counterintuitive behavior is of
fundamental interest because it challenges the conventional wisdom of
equilibrium fluid-solid phase transitions.

\noindent\begin{acknowledgments}

This work was supported by the Office of Basic Energy Sciences,
U.S. Department of Energy, under Grant No. DE-FG02-04-ER46108.
M.C.R. acknolwedges the support of the Natural Sciences and
Engineering Research Council of Canada.

\end{acknowledgments}


\begin{thebibliography}{21}
\bibitem[{Foo()}]{Footnote} \bibinfo{note}{Note that liquid water also undergoes NTE in the temperature range $0^\circ C<T<4^\circ C$ \cite{NTE-Evans}, however, this is not germane to the present work since our focus here is NTE behavior in the crytsalline state.}
\end{thebibliography}
\end{document}